\documentclass{article}
\usepackage[utf8]{inputenc}
\usepackage{authblk}
\usepackage{hyperref}
\usepackage{bm}

\title {Einstein, \emph{Evolution of Knowledge}, and the \emph{Anthropocene}: A Critical Reading of Jürgen Renn's Historiography}
\author{Galina Weinstein}
\affil{\normalsize The Department of Philosophy, University of Haifa, Israel.} 

\begin{document}

\maketitle

\begin{abstract}
This article offers a critical engagement with Jürgen Renn’s historiographical approach, with particular focus on \emph{The Evolution of Knowledge} and \emph{The Einsteinian Revolution} (co-authored with Hanoch Gutfreund). It explores how Renn reinterprets Albert Einstein’s contributions to modern physics, especially special and general relativity, not primarily as the product of individual insight, but as emergent from broader epistemic structures and long-term knowledge systems. The discussion centers on key concepts such as “challenging objects,” “epistemic matrices,” “mental models,” and “borderline problems,” and situates Renn’s framework within broader debates involving Thomas Kuhn, Ludwik Fleck, and Mara Beller. While recognizing the historiographical strengths of Renn’s structuralist approach, the article raises questions about its implications for understanding individual agency, conceptual creativity, and the philosophical dimensions of scientific change. The paper contends that a balanced account of scientific innovation must preserve both the historical embeddedness of knowledge and the originality of conceptual breakthroughs.
\end{abstract}

\tableofcontents

\section{The Evolution of Knowledge}

Jürgen Renn develops a structuralist account of Albert Einstein’s path to special and general relativity in \emph{The Evolution of Knowledge}. Rather than depicting Einstein as a singular revolutionary figure, Renn situates him within a broader historical and cognitive context, emphasizing how scientific innovation emerges from the gradual reorganization of inherited conceptual structures. Einstein’s intellectual development is not portrayed as a moment of isolated genius, but as a response to systemic pressures, conceptual tensions, and the availability of heuristic tools embedded within scientific traditions.

\subsection{Challenging Objects, Mental Models, and Cognitive Disruption}

A central concept in Renn’s framework is “challenging objects," phenomena that disrupt established epistemic expectations and catalyze conceptual restructuring. As a prototypical example, Renn highlights Einstein’s formative childhood experience of observing a compass needle responding to an invisible magnetic force. This event, he argues, violated a deeply embedded assumption in classical mechanics: that motion requires direct physical contact. The compass needle’s movement without visible interaction generated cognitive dissonance, confronting the observer with an epistemic anomaly.

Rather than prompting the abandonment of the concept of motion, such dissonant experiences, according to Renn, initiate a reorganization of internal cognitive structures. The mind adapts by refining its conceptual frameworks—updating or replacing "mental models," those flexible, experience-based representations that guide perception, expectation, and reasoning. In this case, the classical "mental model" of motion was adjusted to account for action at a distance. Although Einstein’s encounter with the compass did not immediately result in theoretical insight, Renn presents it as an early cognitive disruption that sowed the seeds for future conceptual innovation \cite{Ren}.

This example illustrates Renn’s broader claim that scientific understanding evolves through sustained engagement with anomalies that resist assimilation. By integrating insights from cognitive science into historical epistemology, Renn argues that experiences of epistemic friction—like that posed by the compass—drive the long-term restructuring of inherited knowledge. In this light, Einstein’s fascination with the compass becomes emblematic of the dynamic interplay between perception, expectation, and conceptual change that lies at the heart of scientific creativity.

\subsection{Borderline Problems and Conceptual Tensions}

Renn further introduces "borderline problems", which are conflicts between overlapping but incompatible conceptual domains. According to Renn, the so-called “paradigms” such as mechanics and electrodynamics do not undergo a sudden replacement, as in Kuhn’s model. Instead, they overlap in specific regions of application — the "borderline problems." In these zones of conceptual overlap, the seeds of transformation are sown.  

In contrast to Kuhn’s model of abrupt paradigm shifts, Renn proposes a framework of gradual epistemic transformation driven by the emergence of "borderline problems"—zones of conceptual conflict where two distinct systems of knowledge (such as mechanics and electrodynamics) are applied to the same domain but yield incompatible results. Rather than leading to the immediate overthrow of an existing paradigm, these conflicts give rise to "epistemic islands": locally decoupled areas of inquiry where new ideas are explored semi-independently of prevailing frameworks. Over time, these islands may serve as seedbeds for reorganizing accumulated knowledge into a new conceptual architecture. In this view, innovation is not a sudden rupture but a prolonged restructuring process that often spans generations \cite{Ren}. 

A prominent example is the tension between Newtonian gravity and special relativity. Newtonian mechanics assumes instantaneous gravitational action and absolute time, while special relativity introduces a finite speed of light and denies absolute simultaneity. 
This incompatibility created a conceptual crisis that necessitated a reconfiguration of gravitational theory. Renn emphasizes that Einstein's response was not a spontaneous insight but a sustained engagement with these tensions, resulting in the geometric reinterpretation of gravity in general relativity.
The contradiction between Galileo’s principle of free fall and the limitations of special relativity became a "borderline problem," which Einstein approached through the equivalence principle, a "mental model" linking accelerated frames to gravitational fields. This analogy served as an epistemic pivot, allowing Einstein to simulate gravity in new ways \cite{Ren}.

Einstein’s eventual adoption of the metric tensor to represent gravitational potential marked a shift from physical intuition to abstract mathematical formalism. However, Renn emphasizes that this was not a leap by Einstein alone. Instead, it was part of a "protracted," "collective" knowledge integration process, a cyclic negotiation between mathematical representation and physical meaning. With the help of Marcel Grossmann, Einstein pursued a dual strategy: mathematically exploring non-Euclidean geometry while physically grounding the theory in empirical constraints like Newtonian limits. This led to a transitional synthesis, or theoretical "scaffold," culminating in the 1915 field equations.

Yet for Renn, the 1915 publication was just the beginning. The full epistemic integration of general relativity—its "longue durée" extended well into the mid-20th century, involving the emergence of a supportive epistemic community, the resolution of conceptual tensions, and the expansion of the theory into new domains like cosmology and black holes \cite{Ren}.\footnote{The "longue durée" is a historiographical concept, introduced in 1958 by the French historian Fernand Braudel, emphasizing slow-moving, deep structural changes in history over short-term events \cite{Brau}.} 

\subsection{Epistemic Matrices and Conceptual Scaffolding}

To explain how new theories emerge, Renn employs the concept of "epistemic matrices," intermediate cognitive and material resources that support theoretical innovation. These include mathematical tools, analogies, and partial theoretical frameworks. For instance, the Lorentz transformations, devised originally to reconcile Maxwell's equations with the ether theory, served as an epistemic matrix for special relativity. Einstein reinterpreted them as foundational to new kinematics. Similarly, in the development of general relativity, Marcel Grossmann’s introduction of tensor calculus and Riemannian geometry provided Einstein with the formal apparatus to describe gravitational phenomena as spacetime curvature \cite{Ren}.

The 1913 "Entwurf" theory, which Einstein co-authored with Grossmann \cite{EG}, exemplifies what Renn terms a "transitional synthesis" \cite{Ren}. Although it lacked full general covariance and ultimately proved inadequate, it enabled Einstein to refine his mathematical strategies and physical heuristics. This provisional framework facilitated the eventual formulation of the 1915 field equations and illustrates Renn's broader thesis: Scientific progress often proceeds through cyclic and scaffolded conceptual reorganizations rather than abrupt paradigm shifts. Michel Janssen introduced the concept of scaffolding \cite{Jan}.

\subsection{Reinterpreting Inherited Tools}

In Renn’s framework, Einstein’s distinctive contribution is not based on the invention of entirely new concepts, but on his ability to reinterpret existing theoretical tools within novel conceptual frameworks. This is exemplified in Einstein’s treatment of Lorentz’s idea of local time. Originally introduced as a mathematical device to preserve the form of Maxwell’s equations under transformations involving a stationary ether, Lorentz’s “local time” functioned as an auxiliary variable \cite{Lor}. 

According to Renn, Einstein's innovation was to elevate this formal device to a physically meaningful quantity by integrating it into operational procedures involving clock synchronization and light signals. In doing so, Einstein not only adopted Lorentz's formalism ("an auxiliary construct") but fundamentally transformed its ontological implications. Similarly, the metric tensor of Riemannian geometry, originally an abstract construct, was repurposed by Einstein as the central physical entity embodying gravitational effects, thereby bridging mathematical formalism and physical ontology \cite{Ren}.

A notable asymmetry in Renn’s structuralist historiography is the lack of comparable analysis regarding Hendrik Antoon Lorentz’s own epistemic contributions. While Einstein is portrayed as reorganizing inherited materials, particularly Lorentz’s electrodynamics, into a new conceptual order, Renn does not subject Lorentz to the same structuralist scrutiny. What conceptual frameworks did Lorentz inherit, reconfigure, or challenge in constructing his theory of the electron and transformations? This omission creates an uneven historiographical treatment: Lorentz emerges as an originator, Einstein as a reorganizer. If structural reordering is the hallmark of scientific creativity, then Renn’s failure to explore Lorentz’s own intellectual scaffolds weakens the explanatory symmetry of the model. A consistent structuralist account would need to analyze Lorentz’s role in the epistemic lineage with equal rigor, rather than retroactively positioning him as the source of components that Einstein merely rearranged.

\subsection{Slow Reorganization, "Longue Durée", Not Sudden Revolution}

A recurring theme in Renn’s historiography is the rejection of abrupt scientific revolutions. He emphasizes that general relativity, contrary to popular narratives of instantaneous theoretical rupture in 1915, matured over a protracted period through cumulative contributions by a broad scientific community. Although pivotal, the initial formulation of the field equations did not immediately establish the theory as a stable paradigm. Key milestones in this prolonged development included Schwarzschild’s exact solution (1916), Eddington’s empirical confirmation of light deflection (1919), and a period of dormancy through the mid-20th century, followed by a resurgence in the 1960s with developments in black hole physics and gravitational wave theory \cite{Ren}, \cite{BRLR1}, \cite{BRLR2}.

Renn conceptualizes this long arc of development as part of the "longue durée"—a historiographical category denoting gradual structural transformation. He positions general relativity’s eventual consolidation as a paradigmatic case of how epistemic frameworks stabilize only through extensive elaboration, dissemination, and institutional embedding by an epistemic community. Figures such as Friedmann, Weyl, Penrose, and Hawking play integral roles in this process of communal intellectual refinement \cite{Ren}, \cite{GR2}, \cite{BRLR1}.

\subsection{Einstein and the Anthropocene: A Cognitive Analogy}

In \emph{The Evolution of Knowledge,} Jürgen Renn frames the Anthropocene not merely as a geological epoch, but as a historiographical lens through which to reinterpret the long-term development of science. Defined as the era in which human activity has become a planetary force rivaling geological processes, the Anthropocene, for Renn, is the culmination of centuries of accumulated knowledge. It embodies both the empowerment made possible by scientific advancement and the unintended consequences—ecological degradation, inequality, and systemic risk—accompanying that progress.

Renn traces the origins of this epoch through the historical coevolution of knowledge, beginning with the Scientific and Industrial Revolutions and intensifying during the post-WWII period known as the “Great Acceleration.” At the heart of this transformation are "epistemic networks"—self-organizing systems composed of semantic, social, and material layers—that circulate and restructure knowledge. These networks evolve through the resolution of "borderline problems" \cite{Ren}

Einstein’s contributions—particularly the special and general theories of relativity—are presented by Renn as emblematic of this broader process. In this view, special relativity was born at the borderline between Newtonian mechanics and Maxwellian electrodynamics, where conflicting assumptions about space, time, and motion demanded reconciliation. General relativity, in turn, emerged from tensions between special relativity and Newtonian gravity, representing a further synthesis prompted by internal inconsistencies. Importantly, these transformations are not treated as singular acts of genius or abrupt paradigm shifts, but as protracted, multi-generational integrations—historically contingent and shaped as much by institutional and societal conditions as by intellectual insight.

This same logic underpins Renn’s approach to the Anthropocene. The global environmental crisis is not a recent aberration, but the endpoint of a global learning process characterized by the cumulative, though uneven, application of scientific knowledge. Just as general relativity required decades to reach maturity, so too the knowledge needed to respond to the Anthropocene must be seen as emerging from extended, often conflicted historical processes, "the longue durée and incremental phases in the transformation of a system of knowledge" \cite{Ren}. Addressing the Anthropocene demands a reflexive science that recognizes its own embeddedness in social, political, and ecological systems, and accepts responsibility not only for knowledge production but also for its global consequences.

Renn extends his historiographical analysis beyond the domain of physics to propose a broader analogy between Einstein’s conceptual shift in general relativity and the kind of cognitive transformation required to address the challenges of the Anthropocene. Just as Newtonian mechanics had treated spacetime as a static background against which physical phenomena unfold, traditional environmental thinking viewed Earth as a passive setting for human activity. 
In contrast, Einstein’s general theory of relativity reconceptualized spacetime as dynamic and interactive, responsive to the distribution of mass and energy. Renn draws on this shift to argue for a parallel reorientation in planetary thinking, in which humanity recognizes its entanglement within the Earth system \cite{Ren}.
This analogy underpins Renn’s advocacy for a "planetary epistemology," where knowledge systems are not only interdisciplinary but critically attuned to their social and ecological embeddedness. The Einsteinian transformation becomes a model not merely for scientific insight but for epistemic practices capable of addressing complex global problems \cite{Ren}.

\subsection{The Golem, a Copernican reinterpretation and structural concepts}

In \emph{The Evolution of Knowledge}, Renn draws a striking metaphor between science and the Golem of Jewish folklore—a powerful, human-made being fashioned from clay and brought to life through incantation. For Renn, science is not a natural force but a historically contingent construct, shaped by specific cultural, conceptual, and material conditions. Like the Golem, it emerges from its creators' accumulated practices and tensions, emphasizing epistemic matrices and "borderline problems"—frameworks within which scientific knowledge is inherited, challenged, and restructured. Once created, however, science develops its momentum. It evolves through systemic inertia, growing and reshaping itself in ways that exceed the foresight or control of individual actors \cite{Ren}. 

This image parallels Renn’s account of science as a self-organizing system, unfolding over the "longue durée" through processes of structural transformation, of which Einstein’s contributions are interpreted not as singular ruptures, but as exemplary "Copernican processes": reconfigurations of existing elements into a new conceptual order.

Renn uses the Golem metaphor to navigate between two historiographical extremes: on the one hand, the relativist dissolution of science into localized cultural practices, and on the other, the triumphalist narrative of scientific rationality as a linear march of progress. The Golem, powerful, clumsy, and sometimes perilous, offers a middle ground: science as both socially embedded and historically potent but neither infallible nor monolithic.
This metaphor closely aligns with Renn’s framework. In a "Copernican process," scientific innovation proceeds not by discarding the past but by reordering its components, just as the Golem grows by reassembling what it has been given. Einstein did not reject Lorentz or Maxwell; he reinterpreted them. "Borderline problems"—such as the incompatibility of electrodynamics with mechanics—function like faults in the Golem’s programming, prompting systemic reconfiguration. "Epistemic matrices"—"conceptual scaffolds" like tensor calculus or Lorentz transformations—serve as the Golem’s instructions, allowing new behaviors to emerge. And Fleck’s thought "collectives," which Renn draws upon, provides the animating community behind science’s unfolding, like Rabbi Loew summoning the Golem to life. It is collective knowledge and communication that animate the system.

In this light, the Golem is also an implicit metaphor for Einstein himself. The myth of the solitary genius—divinely inspired, transcendent—resembles a theological reading of the Golem: mysterious, unaccountable, and decontextualized. Renn resists this view by embedding Einstein within a systemic process of knowledge evolution. Einstein becomes not a miraculous originator, but an attuned participant—an integrator of existing materials, responsive to conceptual tensions, and guided by shared heuristics. While this structuralist account powerfully demystifies scientific progress, it also risks explaining away the heuristic daring, philosophical judgment, and intellectual initiative that marked Einstein’s most transformative insights.

\subsection{The Limits of Epistemic Evolution}

Renn’s account of Einstein’s work is conceptually rich and grounded in detailed historical analysis \cite{Ren}. However, it arguably underrepresents Einstein’s philosophical agency and decision-making by emphasizing systemic conditions, conceptual inheritances, cognitive scaffolds, and epistemic matrices. Renn risks these by conflating the presence of tools with their creative reconfiguration. The interpretive act of rejecting the ether, relativizing simultaneity, and redefining time in operational terms reflects not only access to certain mathematical resources but a willingness to reorient foundational assumptions.
Renn’s brief mention of Henri Poincaré illustrates this tension. Although Poincaré introduced the term “Lorentz transformations” and anticipated certain features of relativity, he retained a belief in the ether \cite{Po}. He treated the transformations as formal expedients rather than indicators of a new ontology \cite{Ren}. 

By contrast, Einstein eliminated the ether and reformulated the nature of spacetime itself. This constitutes not merely a structural repurposing but a substantive philosophical refoundation that challenges the sufficiency of Renn’s systemic model to capture the originality of Einstein’s theoretical contribution fully.

Nevertheless, Renn does acknowledge key conceptual contributions that reflect Einstein’s intellectual initiative. Einstein’s redefinition of simultaneity, the operational interpretation of time, and the elevation of rods and clocks from experimental tools to foundational theoretical elements are credited as original interpretive acts. These reconfigurations—especially the treatment of Lorentz’s local time as real time—are portrayed not as automatic outcomes of systemic forces, but as creative reinterpretations of inherited structures \cite{Ren}.

\subsection{Kuhn, Paradigms, and Fleck's "thought collective”}

Renn’s engagement with Thomas Kuhn’s paradigm theory is marked by critical distance and selective incorporation. Rather than outright rejecting Kuhn’s model, Renn problematizes its core assumptions. He challenges Kuhn’s conception of scientific revolutions as abrupt, Gestalt-like transformations and instead frames conceptual change as a gradual, structurally embedded process. In this light, revolutions are extended reconfigurations rather than dramatic breaks.

Renn further critiques Kuhn’s model for being overly centered on individual agents and their supposedly sudden insights. Such emphasis, he argues, obscures the underlying systems of knowledge, material culture, and collective epistemic practices that scaffold scientific development. This renders Kuhn’s account insular—too focused on internal conceptual logic and insufficiently attentive to the socio-material conditions that shape scientific thought. The elitism embedded in Kuhn’s conception of "normal science," restricted to specialized communities, also draws scrutiny in Renn’s framework, which favors more pluralistic and distributed knowledge ecologies \cite{Ren}.

In place of Kuhn, Renn aligns more closely with Ludwik Fleck, whose notion of “thought collectives” and emphasis on resistance from reality resonate with Renn’s categories, such as “challenging objects.” Fleck’s attention to the co-evolution of epistemic styles and material contexts allows Renn to develop a historical epistemology that avoids Kuhn’s Cold War–era idealism and individualism. Thus, relativity is not treated as a Kuhnian paradigm shift but as a "longue-durée" reorganization of knowledge \cite{Ren}, \cite{Fl-1}, \cite{Fl-2}, \cite{Kun}. 

Special relativity emerged from synchrony procedures, Maxwellian electrodynamics, and Lorentz transformations. The "longue durée" refers to the gradual reorganization of scientific understanding and institutional acceptance of the theory after 1905, through reinterpretations by Minkowski, Planck, von Laue, and others, and the eventual embedding of relativistic principles in experimental practice. General relativity, in turn, requires communal elaboration across decades before crystallizing into a stable framework. In this model, Einstein is not a paradigm-maker but a participant in an evolving epistemic network. Paradigms are recognized only in hindsight and retrospectively constructed through stabilization (if they exist).

\subsection{Conclusion: Agency, Structure, and the Boundaries of Innovation}

Renn interprets the emergence of special relativity as a case of knowledge transformation at the boundary zones of classical physics. He implements his key concepts, “borderline problems,” “epistemic islands,” “mental models,” and “knowledge integration, ” to show how Einstein’s 1905 breakthrough emerged from tensions within the established domains of mechanics and electrodynamics. 

At the turn of the 20th century, classical physics was compartmentalized into subfields: mechanics, thermodynamics, and electrodynamics. However, developments in these domains produced “borderline problems.” One such tension arose in the electrodynamics of moving bodies, where contradictions between the principle of relativity in mechanics and the ether-based interpretation of light propagation in electrodynamics formed a conceptual fault line. Lorentz had constructed an increasingly elaborate theory to resolve these tensions, including \emph{ad hoc} hypotheses like length contraction and local time \cite{Ren}. 

Renn sees Lorentz’s work as a significant “epistemic effort,” which became conceptually strained and dependent on questionable physical assumptions (e.g., the stationary ether). Einstein perceived that this field had become an “epistemic island,” a region of physics isolated from the rest of classical mechanics due to internal contradictions. What distinguishes Einstein’s approach is his decision to abandon the ether and interpret the Lorentz transformations not as auxiliary mathematical tools but as physically meaningful. By combining two levels of knowledge, theoretical (electrodynamics) and practical (time measurement with rods and clocks), Einstein transformed the conceptual foundations of physics. He introduced two postulates: the constancy of the speed of light and the principle of relativity, and from these, rederived the Lorentz transformations as consequences, not assumptions. This move inverted the logic of Lorentz’s theory and created a new kinematical framework grounded in a redefinition of simultaneity \cite{Ren}. 

This process exemplifies Renn’s metaphor of “exploring the horizon” of classical physics, identifying where existing frameworks fall short, and developing new structures through reflective abstraction. Einstein’s breakthrough thus becomes a case study in how “borderline problems” catalyze the emergence of new knowledge systems. The resulting special theory of relativity initially remained an “epistemic island,” a theory valid only for inertial frames. Still, it gradually transformed into a central pillar of modern physics through a conflict-laden process of integration. This process echoes Renn’s emphasis on the "protracted" and "collective" nature of conceptual change, which Renn refers to as the "longue durée" \cite{Ren}. 

Renn’s reading of Einstein offers a compelling reorientation of how major scientific transformations occur. It situates conceptual change within systemic dynamics, emphasizing the importance of inherited tools, community elaboration, and socio-cognitive scaffolding. Through mechanisms and metaphors, such as "challenging objects," "borderline problems," "epistemic matrices," "mental models," and "longue durée,"  Renn reframes the genesis of special and general relativity as cases of structured innovation rather than singular invention.

Nevertheless, this perspective comes at a cost. Its commitment to structural coherence downplays the heuristic imagination and philosophical discretion that shaped Einstein’s trajectory. Ultimately, Renn places Einstein’s work within his broader model of knowledge evolution, where scientific innovation is seen less as a product of isolated genius and more as the outcome of tensions within systems of knowledge, shaped by historical, institutional, and epistemic conditions. Einstein’s insight was pivotal, but it is interpreted as part of a broader historical dynamic, not as a \emph{sui generis} revolution. Einstein’s abandonment of the ether, reconceptualization of simultaneity, and embracing general covariance are framed not as revolutions of thought but as adaptations within a systemic logic. As such, his role becomes illustrative rather than exceptional.

The analytical strength of Renn’s model lies in its capacity to demystify scientific change. However, this same strength may obscure the particularity of Einstein’s interventions—the intellectual risks taken and the conceptual daring exercised. In doing so, the model flattens creativity into emergence and blurs the distinction between structural possibility and philosophical initiative. Renn’s historiography, while rich in insight, ultimately raises a broader question: can structuralism fully account for scientific novelty without eroding the very agency it seeks to contextualize?

\section{The Einsteinian Revolution}

\subsection{Reframing Originality within Epistemic Structures}
Building on the conceptual architecture introduced in \emph{The Evolution of Knowledge}, \emph{The Einsteinian Revolution} extends Jürgen Renn’s structuralist methodology to systematically reinterpret Einstein’s role in modern physics. Central to this project is a deliberate shift away from the figure of Einstein as a solitary genius and toward a depiction of him as a participant in a broader, evolving epistemic system \cite{GR4}.

The authors explicitly aim to “dispel the popular myth that Einstein, the unconventional scientific genius, instigated an overwhelming scientific revolution through pure thought alone” \cite{GR4}. In its place, they offer a model in which Einstein’s contributions are seen as responses to long-standing epistemic tensions and boundary problems, particularly the frictions between electrodynamics, thermodynamics, and classical mechanics. Einstein’s originality, in this framing, is not rejected but redefined as an emergent property of preexisting conceptual constraints.

This reinterpretation embeds Einstein’s thinking in a continuum of developments stretching from Galileo and Newton to Maxwell and Lorentz. In this view, scientific advances are not abrupt ruptures born of intuition or inspiration but structured shifts within the deep currents of conceptual evolution. Einstein’s creativity becomes a function of the systemic intelligibility of the problems he addressed—a process guided by inherited cognitive "scaffolding" rather than spontaneous insight.

Renn’s key metaphor, the “Copernican reinterpretation,” encapsulates this method: as Copernicus reorganized astronomical models without positing new celestial entities, Einstein is said to have reorganized classical physics without introducing fundamentally novel principles. The innovation, then, lies in conceptual realignment, not metaphysical disruption. As the authors put it, “The substance of Einstein’s work was not new but rather was the result of an accumulation of knowledge over centuries; it was his conceptual organization that was new” \cite{GR4}.

This model treats epistemic constraints such as general covariance, energy-momentum conservation, and the correspondence principle as guiding structures that shaped Einstein’s reasoning. They served as cognitive boundaries within which his theories crystallized. This “top-down and bottom-up assimilation” of knowledge positions Einstein as a synthesizer of circulating ideas rather than their originator \cite{GR4}. His genius is recoded as epistemic integration.

Yet this structuralist apparatus, while compelling in its depiction of continuity and transformation, raises historiographical stakes. By emphasizing the role of systemic pressures and inherited matrices, the model risks minimizing subjectivity, masking heuristic innovation, and neutralizing the daring philosophical choices that defined Einstein’s path. His rethinking of simultaneity, the ontological status of spacetime, and the rejection of absolute frames are not merely redistributions of existing content—they are interpretive leaps. Although Gutfreund and Renn stop short of erasing Einstein’s agency, they embed it so deeply in systemic logic that its independence is difficult to retrieve.

\subsection{Demythologizing Through Structure}

Renn and Gutfreund's structuralist interpretation intensifies in their treatment of Einstein's engagement with classical physics. They argue that Einstein's significant theoretical contributions arose in response to Renn's preexisting "borderline problems" that created instabilities within the conceptual edifice of classical mechanics and electrodynamics. These tensions guided Einstein toward reconfiguring existing theoretical components rather than initiating a radical departure from them \cite{GR4}. Consequently, Einstein appears less as a theorist who generates fundamentally new frameworks and more as one who synthesizes and reorders inherited knowledge. His theories become epistemic consolidations, not creative ruptures.

This framing applies both to special and general relativity. Special relativity is described not as a foundational break, but as a restructuring of tensions at the boundary of nineteenth-century physics, particularly among Maxwell's equations, Lorentz transformations, and Machian critiques. General relativity, in turn, is presented as the systemic outcome of reconciling Newtonian gravitation with relativistic kinematics under the guidance of high-level heuristic principles. These include general covariance, energy conservation, and the Newtonian limit \cite{GR4}. By foregrounding these structural pressures, the authors position Einstein less as a conceptual innovator and more as a synthesizer constrained by the trajectory of knowledge evolution.

\subsection{Mental Models as Thought Experiments in Einstein’s Case}

This structuralist lens becomes most visible in reinterpreting Einstein's thought experiments. Once celebrated for their imaginative and philosophical depth, experiments like the lightning train or the rotating disk are reframed in \emph{The Einsteinian Revolution} as heuristic devices—“mental models”—rather than expressions of philosophical creativity. The authors explicitly downplay psychological or intuitive accounts of Einstein's imagination, stating that such perspectives are “not very helpful” for understanding the development of scientific knowledge. In doing so, they emphasize cognitive "scaffolding" over individual insight \cite{GR4}.

Yet Einstein’s thought experiments can be fruitfully understood as mental models in a richer sense: internal, flexible representations that bridge concrete experience and abstract reasoning. Rooted in cognitive science and influenced by Gestalt psychology, mental models enable individuals to manage incomplete information, test hypothetical scenarios, and reorganize conceptual structures without discarding the entire framework. In Einstein’s case, canonical examples such as the train-lightning platform scenario or the moving clock illustrate how familiar experiential domains—observers, rods, and synchronized clocks—served as intuitive scaffolds for navigating the tensions between Galilean relativity and Maxwellian electrodynamics. These models enabled Einstein to probe the relativity of simultaneity and reconceptualize time as a relational and operational quantity. Far from being pedagogical afterthoughts, such thought experiments functioned as essential discovery tools: they integrated empirical reasoning with mathematical formalism (e.g., Lorentz transformations), facilitating structural reorganization within the conceptual matrix of classical physics.

\subsection{Three Epistemic Challenges}

Gutfreund and Renn outline three epistemic challenges that purportedly define the path to general relativity \cite{GR4}: 

1. \emph{The Challenge of Missing Knowledge:} Gutfreund and Renn ask how Einstein could construct general relativity without clear empirical anomalies demanding a new gravitational theory. Except for the perihelion anomaly of Mercury, traditional physics still adequately explained most known phenomena when Einstein began his search. Gravitational waves, black holes, and other now-fundamental implications of general relativity had not yet been discovered. To explain the remarkable precocity of Einstein’s theory, the authors point to the long-term accumulation of scientific knowledge across centuries. They argue that the conceptual and mathematical infrastructure—ranging from the laws of planetary motion to non-Euclidean geometry and the development of differential calculus—formed a shared epistemic reservoir that Einstein could tap into and reorganize. Even before general relativity was completed, figures like Schwarzschild had already begun applying this knowledge to astronomical problems \cite{GR4}.

In this account, Einstein’s achievement was not the creation of entirely new content but the strategic recombination of inherited resources. His particular cognitive perspective, shaped by classical physics and philosophical reflection, distinguished his success, which enabled a Copernican-style reconfiguration of accumulated insights. Thus, what appears as revolutionary genius is reframed as the outcome of navigating shared knowledge differently. The novelty lay not in the tools but in Einstein’s ability to deploy them toward a new synthesis. 

However, Gutfreund and Renn's framing reduces Einstein’s theoretical leap to a logical unfolding of inherited knowledge, sidestepping the imaginative synthesis required to connect these elements in an unforeseen way. That is a diminishing move, whether intended or not. This interpretation misconstrues the nature of Einstein’s theoretical leap. Possessing the conceptual “Lego bricks” of the time does not imply that assembling them into a revolutionary theory was inevitable. The true insight lay in seeing a structure no one else imagined possible. By attributing the achievement to the slow accumulation of background knowledge, the authors blur the distinction between the availability of tools and the originality of vision required to unify them.

2. \emph{The Challenge of Delusive Heuristics:} Gutfreund and Renn frame Einstein’s path to the correct field equations of general relativity (in \cite{Ein15-1}) as an illustrative case of “the challenge of delusive heuristics.” They focus on the puzzling trajectory in which Einstein first arrived at the correct form of the equations in 1912–1913 (while working with the Zurich Notebook), only to abandon them in favor of the "Entwurf" theory, and finally returned to the correct form in 1915. This apparent regression is interpreted not as a failure of reasoning but as a symptom of the inherited mental models from classical physics that shaped and constrained Einstein’s thinking.

According to the authors, Einstein’s heuristics—principles such as the equivalence principle, conservation laws, correspondence principle (Newtonian limit), and covariance—were not arbitrary tools but deeply rooted in classical field theory's epistemic architecture, especially as demonstrated by Galileo, Newton, and Lorentz. These heuristics enabled Einstein to formulate meaningful criteria for a gravitational field equation, but they also led to internal tensions that could not be resolved through intuition alone. Einstein’s difficulty lay not in the absence of knowledge, but in the abundance of it. Too many inherited assumptions about how a field should behave, be sourced, and conform to classical mechanics and field theory had to be reconciled.

In this account, Einstein's heuristics are both empowering and misleading: they arise from the deeply internalized knowledge structures of classical physics and Lorentzian field models, but must be tested, refined, and sometimes abandoned to yield truly novel theory. Einstein's vacillation between physical and mathematical strategies is interpreted as a dynamic interplay between alternative pathways of reorganizing knowledge. The final success—identifying the Einstein tensor and its link to the stress-energy tensor—is portrayed not as a sudden inspiration but as the product of a drawn-out process in which Einstein systematically recalibrated default assumptions and heuristic tools \cite{GR4}.

The more profound implication is epistemological: Einstein did not transcend his intellectual inheritance so much as reorganize it. His genius lay not outside the system but in his capacity to reconfigure its architecture from within. This interpretation aligns with Renn’s broader thesis that scientific revolutions are not abrupt paradigm shifts, but protracted transformations of inherited knowledge systems. In this framing, Einstein’s seeming detours and returns become moments in a nonlinear evolutionary process rather than signs of individual brilliance or failure.

The challenge of delusive heuristics, as framed by Gutfreund and Renn, recasts Einstein’s conceptual struggles not as signs of creative dynamism but as symptoms of epistemic entrapment. His oscillations between candidate field equations—from the Ricci tensor in the Zurich Notebook to the flawed "Entwurf" field equations and back—are interpreted as consequences of inherited classical principles: Newtonian correspondence principle, the equivalence principle rooted in Galileo’s law of free fall, and the conservation laws of classical mechanics. Rather than viewing these as heuristic tools Einstein critically engaged with, the authors portray them as systemic constraints that dictated his missteps. In this narrative, Einstein’s agency is displaced by the inertia of classical physics; discovery becomes less an act of individual breakthrough than the unfolding of an evolutionary logic. Trial and error, typically the mark of scientific creativity, is reframed as delusion born from structural inheritance. Thus, Einstein is subtly pushed into the background—not the agent who overcomes barriers, but the figure through whom the past continues to speak.

3. \emph{The challenge of discontinuous progress:} Gutfreund and Renn identify a third epistemic challenge in Einstein’s development of general relativity: the challenge of discontinuous progress. This challenge centers on the tension between the theoretical foundations Einstein inherited, rooted in classical and special relativistic physics, and the radically nonclassical outcomes of his work, particularly the general theory of relativity’s treatment of spacetime as dynamic and geometrized. Such a leap, they argue, cannot be understood merely as an extension of existing knowledge but must be seen as a transformation of the knowledge system itself.

Einstein’s work is thus portrayed not as a linear enhancement of Lorentzian electrodynamics or classical mechanics, but as a reflective process in which preexisting heuristic principles—such as the equivalence principle and the conservation principle—were assimilated, tested, and ultimately reconfigured. They emphasize that this transformation was not solely a top-down application of abstract cognitive structures, but a dynamic interplay: Einstein modified and reinterpreted these structures through experience and mathematical experimentation. This process, which they describe as the “reflection phase,” was the turning point in reorganizing classical knowledge into a fundamentally new system \cite{GR4}.

By framing Einstein’s innovation in this way, the authors position general relativity not as an abrupt Kuhnian paradigm shift but as a protracted and cyclic reworking of conceptual resources. In this view, the leap from classical assumptions to a geometric theory of gravitation required both the retention and reinterpretation of existing principles, culminating in a theory that, while historically situated, fundamentally broke with its origins. This non-linear evolution, grounded in feedback between experience, formalism, and conceptual reflection, renders general relativity a uniquely discontinuous form of scientific progress.

As presented by Gutfreund and Renn, the challenge of discontinuous progress downplays the ontological rupture introduced by Einstein, namely, the replacement of force with geometry and the abandonment of absolute space and time. Instead of recognizing this as a decisive conceptual break, they recast it as a smooth transformation within a broader epistemic continuum. What should be understood as a radical shift in the metaphysical foundations of physics is reframed as the self-adjustment of knowledge systems. In their account, Einstein does not leap beyond classical physics—he converges with it. The profound conceptual reorientation at the heart of general relativity is thus absorbed into a historical narrative that privileges continuity over rupture, flattening Einstein’s originality into a systemic inevitability.

\subsection{Rebel or Relic? Einstein and the Narratives of Epistemic Evolution}

The structuralist model’s interpretive clarity comes at a narrative cost. In translating Einstein’s breakthroughs into the language of epistemic frameworks and conceptual inheritance, \emph{The Einsteinian Revolution} subtly bureaucratizes his image. His theories emerge not as acts of intellectual insurgency but as the systematized culmination of an evolving knowledge order. What gets sidelined is the visceral charge of rebellion that long animated Einstein’s cultural and scientific persona.

This tonal flattening is evident in the book’s early biographical framing. Einstein is portrayed not as a provocative thinker drawn to paradox and irony, but as a dutiful student who disliked rote learning and proceeded by academic perseverance \cite{GR4}. Missing is the irreverent adolescent who penned satirical dialogues mocking philistine teachers, the young patent clerk who devised radical thought experiments while defying academic conventions, and the mature scientist who toyed with limericks and philosophized about God and dice.

This is not an incidental omission. It flows directly from the authors’ structuralist lens, foregrounding institutional dynamics, material constraints, and long-term epistemic shifts over personal vision and eccentricity. The Einstein who emerges is a functionary of cognitive evolution—a nodal point in a self-organizing system, rather than a visionary who challenged prevailing frameworks. Intellectual independence, playfulness, defiance, and aesthetic judgment recede from view.

The result is a form of biographical compression that substitutes contingency for insight and continuity for rupture. Traits that once marked Einstein as singular—his heuristic audacity, metaphysical daring, and capacity to invert assumptions—are absorbed into the quiet rhythm of evolutionary development. The drama of theory-building is reframed as the administrative task of conceptual coordination.

Ironically, in seeking to puncture the myth of the isolated genius, the authors risk constructing a mirror myth: that of epistemic inevitability, where the scientist becomes incidental to the unfolding of structural logic. The consequence is not merely analytical but emotional. What fades is the unruly brilliance that made Einstein not just a figure within the history of knowledge, but a rebel who redefined its boundaries.

\subsection{Historiographical Asymmetry: Lorentz Elevated, Einstein Subsumed}

One of the most revealing historiographical patterns in \emph{The Einsteinian Revolution} is the asymmetric treatment of Hendrik Lorentz and Albert Einstein. This asymmetry runs deep. While Gutfreund and Renn apply their structuralist framework with rigor and granularity to Einstein, dissolving his agency into the epistemic "scaffolding" of knowledge evolution, Lorentz is portrayed with notable latitude. Despite his theoretical reliance on ether constructs and \emph{ad hoc} mechanisms, Lorentz is granted conceptual originality and historiographical leniency. Einstein, by contrast, is treated as a reorganizer of inherited tools, a participant in epistemic cycles rather than a revolutionary thinker.

In their account of special relativity, the authors credit Einstein with identifying the contradiction between Galilean mechanics and Maxwellian electrodynamics—a paradigmatic "borderline problem." Yet, they emphasize that his resolution emerged from reorganizing existing structures: the Lorentz transformations, clock synchronization conventions, and inherited heuristic assumptions. Einstein is lauded not for introducing fundamentally new concepts, but for reinterpreting old ones. The Lorentz model becomes a "scaffold" and constraint, an epistemic apparatus that Einstein worked through. However, the interpretive generosity stops with Lorentz: his theory \cite{Lor}, despite being rooted in a discredited ether paradigm and relying on length contraction and local time as fixes, is described as "extraordinarily successful" and "systematic" \cite{GR4}

Lorentz is allowed to be original despite his empirical shortcomings. Einstein, by contrast, is depicted as derivative even in his most revolutionary moments. His reinterpretation of simultaneity, space, and time is recoded as the systematized culmination of an epistemically overdetermined framework. This selective application of the structuralist lens—rigorously applied to Einstein, leniently applied to Lorentz—results in a historiographical inversion: Lorentz is the originator, Einstein the synthesizer.

The asymmetry deepens in the treatment of general relativity. Here again, Einstein's return to the Ricci tensor in 1915 \cite{Ein15-1}, after the "Entwurf" period (\cite{EG}, \cite{Ein14}), is framed as a typical "Copernican process"—a cognitive cycle of reinterpretation rather than a moment of radical departure. His collaboration with Grossmann is emphasized to underscore his dependency on existing mathematical tools. Even his November 1915 identification of the field equations (\cite{Ein15-1}, \cite{Ein15-3}) is presented as a systemic convergence, not a philosophical rupture. Meanwhile, Lorentz’s influence (the "Lorentz model") lingers in the background unexamined, implicitly cast as a foundational model rather than a figure embedded in his own network of dependencies—Fresnel, Maxwell, Hertz, Poincaré.

The narrative further credits the \emph{Einstein–Besso manuscript} of 1913 (\cite{CPAE4}, Doc. 14) as a key heuristic precursor to Einstein’s successful derivation of Mercury’s perihelion advance in 1915 \cite{Ein15-2}. While the manuscript was indeed reused, Gutfreund and Renn risk conflating structural continuity with conceptual innovation. They treat this earlier manuscript as an enabling "scaffold," downplaying the novel insights in the final formulation, such as Einstein’s use of Christoffel symbols and his refinement of the Newtonian limit. The implicit message is that Einstein’s achievement was epistemically encoded from the outset.

Yet consistency would demand that Lorentz's intellectual debts be subjected to the same structuralist scrutiny. His electrodynamics did not arise in a vacuum: it was indebted to Maxwell’s equations, Fresnel’s ether theory, and the prevailing mechanistic worldviews of 19th-century physics. His own conceptual shifts—such as local time—were responses to experimental anomalies, shaped by contemporary scientific currents. However, Gutfreund and Renn treat Lorentz as an autonomous theoretical agent while Einstein is described as a node in a self-organizing epistemic network.

This imbalance produces an ironic historiographical outcome. In a book dedicated to demythologizing the lone genius, the authors end up preserving the myth—just not for Einstein. Lorentz, whose theory ultimately failed to satisfy the principle of relativity, is cast as a visionary thinker; Einstein, who redefined the foundations of physics, is cast as a well-positioned integrator. The protagonist of \emph{The Einsteinian Revolution} is structurally marginalized in his own narrative.

The deeper flaw lies in the structuralist method itself: it flattens agency selectively. Some figures are spared—granted conceptual autonomy and contextual latitude. Others, like Einstein, are epistemically neutralized. His originality is bureaucratized; his leaps of imagination reframed as recursive permutations of inherited structures. This asymmetry is not merely a rhetorical problem—it undermines the credibility of the historiographical framework. If the evolution of knowledge is to be understood as a collective, cumulative process, then its application must be symmetrical. Otherwise, the result is a historiographical distortion: one in which "scaffolding" is emphasized for some and erased for others, and in which the very thinker who revolutionized the cosmos is rendered an accessory to epistemic inevitability.

\subsection{The Fate of Free Creations of the Human Mind}

The structural irony at the heart of \emph{The Einsteinian Revolution} is that while the book centers explicitly on Einstein as its main protagonist, its historiographical strategy systematically diminishes him. Other historical figures—such as Lorentz, Poincaré, or Grossmann—appear only fleetingly, often in supporting roles, yet are granted more conceptual autonomy than Einstein himself. The narrative framework does not expend its analytic depth on them, but on the scaffolding that allegedly shaped Einstein’s thought. Thus, while Einstein is nominally foregrounded, he is functionally displaced. His philosophical judgment, heuristic risk-taking, and creative agency are subsumed into the very structure that claims to explain him. This paradox results in a protagonist who is always present but never fully empowered, reduced to the status of a cognitive relay in a larger system whose contours are granted more agency than the thinker at its center.

Gutfreund and Renn acknowledge Einstein’s famous dictum that scientific theories are “free creations of the human mind,” yet this recognition is subtly subordinated to their broader structuralist narrative. While they explore Einstein’s thought processes through the lens of Gestalt psychology, "mental models," and epistemic constraints, they ultimately embed his creativity within a systemic Copernican framework of knowledge restructuring. Conceptual transformations are depicted as adaptive reorganizations rather than radical departures. Einstein’s imaginative insights—his thought experiments, his redefinition of simultaneity, and his epistemological daring—are recast as instances of structural reconfiguration rather than autonomous innovation. The rhetorical inclusion of “free creations” thus coexists with, but is ultimately contained by, a historiographical logic that privileges continuity over rupture and scaffolding over singular conceptual leaps.
In the end \emph{The Einsteinian Revolution} leaves us with a paradoxical tale, one in which the man whose theories redefined the cosmos is rendered a passive conduit of epistemic currents, his "free creations of the human mind" recast as inevitable permutations of inherited structures, and his singular light dimmed beneath the very scaffolding he once transcended.

This tension becomes even more evident in Gutfreund and Renn’s portrayal of Einstein’s intellectual persona. While they acknowledge his resistance to disciplinary boundaries and emphasize his capacity to synthesize insights across fragmented domains, this celebration of cognitive freedom is undercut by the systemic logic of their narrative. Einstein is praised for transcending institutional constraints, for working on the margins of academic science, and for unifying disparate strands of knowledge—yet these very attributes are ultimately folded back into the long-term evolution of knowledge systems. Their structuralist framework absorbs Einstein’s originality into the gradual unfolding of epistemic reconfiguration. The “free creation” becomes not a spark of innovation, but an emergent property of the system. Thus, even the portrayal of Einstein as an outsider—iconoclastic, irreverent, and independent—is paradoxically used to reinforce the continuity of systemic transformation. His creative ruptures are presented not as disruptions but as the natural convergence points of inherited tensions. In seeking to historicize Einstein, Gutfreund and Renn succeed in contextualizing him—but at the cost of diminishing the radical freedom that once defined his legacy.

\subsection{Beller's Dialogism Recontextualized}

The tension within Gutfreund and Renn's historiographical framework becomes especially apparent in their invocation of Mara Beller’s theory of dialogism \cite{Bel}. In \emph{The Einsteinian Revolution}, dialogism is presented as a historiographical alternative to Kuhn’s rigid model of paradigms and incommensurability. Beller’s critique of Kuhn is enlisted to support the authors’ emphasis on distributed epistemic activity and the cumulative transformation of knowledge through networked interactions. Within this framing, Einstein’s achievements are depicted not as the product of philosophical daring or intellectual rupture, but as outcomes of embedded conceptual dialogues—exchanges within a collective structure rather than acts of visionary originality.
Beller’s dialogism allows Gutfreund and Renn to claim that Einstein’s theory of relativity was not a revolution but part of a gradual, networked transformation. 
Einstein was influenced by many contributors (e.g., Grossmann, Besso, Minkowski), and his work was scaffolded on older frameworks \cite{GR4}.

However, this deployment of Beller’s work is selective and arguably distorts her intent. In \emph{Quantum Dialogue} (1999), Beller introduced dialogism not as a mechanism for dissolving creativity into systems theory, but as a humanistic and pluralistic account of how conceptual innovation emerges through tensions, miscommunications, rivalries, and interpretive struggle. Her project aimed to restore intellectual agency and philosophical imagination to the historical record, countering the flattening effects of both Kuhnian incommensurability and institutional determinism. Her dialogical analysis of figures such as Einstein, Bohr, and Heisenberg foregrounded their imaginative labor and conceptual risk-taking, not their passive assimilation into epistemic scaffolds \cite{Bel}.

By contrast, Gutfreund and Renn instrumentalize dialogism to legitimize a structuralist historiography that effectively neutralizes Einstein’s agency. In their rendering, dialogue is no longer a source of ideational novelty but a conduit for integrating and reorganizing preexisting knowledge. These reframings transform dialogism from a celebration of situated creativity into a theoretical tool for justifying the displacement of individual authorship. The irony is that Beller’s theory, originally meant to resist the depersonalization of science, is here appropriated to support precisely that outcome.
Even more striking is that this reinterpretation bypasses existing scholarship that extends Beller’s dialogism to the history of Einstein’s work, particularly studies that have employed her framework to interpret the development of Einstein’s gravitational theory as emerging through dialogical exchanges with figures such as Gunnar Nordström, David Hilbert, and others.

From the standpoint of a scholar trained by Beller, the above structural flattening is especially jarring. Beller’s dialogism sought to reintegrate intellectual agency into the historiography of science. Her goal was to highlight the human dimension of conceptual transformation—the imaginative tensions, philosophical choices, and dialogical struggles that define scientific creativity. \emph{The Einsteinian Revolution}, by contrast, neutralizes these dimensions. It renders Einstein passive and procedural, reducing his originality to systemic adaptation.

This deterministic rendering culminates in a narrative of epistemic convergence. The authors argue that Einstein’s contributions were integrations rather than inventions—inevitable products of conceptual trajectories long underway. His intuition, thought experiments, and aesthetic sensibilities are characterized as distractions from the real drivers of knowledge transformation: structural alignment and heuristic recombination. Yet this account leaves a critical historiographical question unresolved: Why Einstein, and why then? If the groundwork was universally available, why did only he act upon it?

\subsection{Renn’s Intensification of Structuralism}

In \emph{The Evolution of Knowledge}, Renn presents a nuanced structuralist account. Still, he acknowledges the depth of Einstein’s conceptual innovation, especially his epistemic daring to redefine simultaneity and space-time. Einstein is situated within a complex epistemic network but has also been shown to challenge and transcend its constraints. Renn allows Einstein’s heuristics, exploratory reasoning, and conceptual transformation to shine through, even as he embeds them in a broader framework. The metaphors of “epistemic islands,” “scaffolding,” and “knowledge integration” still function structurally, but they do not erase Einstein’s creative role. Instead, they illustrate how Einstein navigated, reorganized, and restructured inherited knowledge systems.

In \emph{The Einsteinian Revolution}, Gutfreund and Renn adopt a more flattened, depersonalized structuralism. The “Copernican process” metaphor dominates: Einstein is not portrayed as a revolutionary, but as a reorganizer of inherited frameworks—someone who revisits discarded ideas (like the Ricci tensor) and reframes them within matured epistemic matrices. His breakthroughs are reinterpreted not as intuitive leaps or singular acts of genius, but as systemic realignments that were structurally bound to happen. His originality is dissolved into long-term cognitive cycles, while other figures, like Lorentz, are depicted as visionary despite conceptual limitations.
Several developments explain the increasingly deterministic position adopted in \emph{The Einsteinian Revolution}:

\subsubsection{1. From Structural Evolution to Deterministic Systems}

In \emph{The Evolution of Knowledge}, Renn offered a pluralistic account of knowledge formation that integrated individual cognition with broader material and epistemic systems. Although structural constraints already moderated individual creativity, the narrative left interpretive space for heuristic innovation, conceptual reorientation, and epistemic agency. Einstein was still treated as a navigational figure, one whose insight catalyzed transformation.

In \emph{The Einsteinian Revolution}, this space contracts. Einstein is no longer a navigator but a conduit, assimilating ideas shaped by systemic forces; his decisions and conceptual breakthroughs are portrayed as structurally overdetermined. Rather than responding inventively to philosophical tensions, he is shown assimilating inherited ideas according to pre-existing heuristic logics. What was once a nuanced theory of evolving knowledge becomes, retroactively, a justification of systemic inevitability. The shift marks a transition from structural evolution to structural determinism.

\subsubsection{2. The Alliance with Hanoch Gutfreund}

Renn maintains a reflective structuralist position in his solo and earlier collaborative works. He emphasizes the systemic nature of knowledge change but still leaves space for individual agency, heuristic invention, and conceptual daring \cite{RR}. Einstein is framed as someone who recognized "borderline problems," drew on inherited structures, and imposed a philosophical reinterpretation—a Copernican act not only in method but in vision. By contrast, Gutfreund and Renn adopt a more rigid and deterministic tone in \emph{The Einsteinian Revolution}. Einstein’s creativity is compressed into a function of epistemic convergence. Dialogue and heuristic boldness are absorbed into structural inevitability. Even imaginative leaps—such as the Ricci-based 1912 equations or the 1915 Mercury paper \cite{Ein15-2}—are recoded as the systematized culmination of inherited "scaffolds."

The collaboration with Hanoch Gutfreund appears to have contributed to Renn's structuralism's rhetorical and historiographical hardening. Gutfreund, long known as the steward of Einstein’s legacy, had previously celebrated Einstein’s originality, philosophical daring, and the emotional arc of discovery. In his 2019 preface to the facsimile edition of Einstein’s 1916 general relativity manuscript, Gutfreund characterized Einstein’s journey as a “convoluted intellectual odyssey,” described its resolution as a “triumphant end,” and called the work itself the “Magna Carta of modern physics.” He vividly emphasized Einstein’s “brilliant insights,” false starts, and eventual conceptual leap from force to geometry \cite{Gut}. This rhetorical framing highlighted Einstein’s agency, resilience, and visionary imagination.

Yet in \emph{The Einsteinian Revolution,} Gutfreund co-signs a narrative that markedly diminishes this vision. Rather than mediating Renn’s structuralist interpretation, Gutfreund’s presence amplifies it, providing institutional authority and symbolic endorsement for a historiographical framework that systematically neutralizes Einstein’s creative autonomy. The philosophical daring once foregrounded is now reframed as the epistemically overdetermined outcome of structural constraints.

This shift introduces a striking paradox. A figure once associated with commemorating Einstein’s singular intellect now participates in a narrative that portrays him less as a conceptual originator and more as a node within a knowledge system reorganizing itself. Einstein’s major breakthroughs are presented not as intuitive syntheses or philosophical ruptures, but as logical convergences of inherited structures. The myth of genius is not merely dismantled—it is supplanted by a model of structural determinism in which the historical actor is effectively dissolved.

This reframing is reinforced by the repurposing of material from Gutfreund and Renn’s earlier collaborations, such as \emph{The Road to Relativity} \cite{GR1}, \emph{The Formative Years of Relativity} \cite{GR2}, and \emph{Einstein on Einstein} \cite{GR3}. In those works, Einstein’s interpretive struggles, evolving heuristics, and conceptual tensions were treated with analytic depth and narrative nuance. In \emph{The Einsteinian Revolution}, these episodes are recontextualized as illustrations of systemic convergence. The result is not only a historiographical realignment but a philosophical flattening: the thinker once celebrated for transcending frameworks is now folded into one.

What emerges from this alliance is a synthetic account that appears coherent and layered but rests on a displacement of conceptual agency. Einstein has become less of a philosophical innovator and more of an epistemic relay. The very achievements once portrayed as “free creations of the human mind” are reinterpreted as structurally encoded transformations. Thus, the alliance between Gutfreund and Renn culminates not only in a consolidation of prior work but in a historiographical closure that recasts Einstein himself, less as a creator of new knowledge architectures than as a predictable expression of their evolution.

\subsubsection{3. Reaction Against Hero Worship and Genius Narratives}

Recent historiography of science has seen a pronounced shift away from the lone genius myth, a move motivated by the desire to correct distortions introduced by hagiographic narratives. This corrective trend is grounded in a legitimate concern: romanticized portraits of solitary figures such as Newton or Einstein can obscure the complex, collective, and contextual nature of scientific development. However, \emph{The Einsteinian Revolution} exemplifies a tendency to overcorrect. In this work, Gutfreund and Renn not only dismantle the Einstein myth, but they also replace it with an equally totalizing narrative grounded in structural determinism.

Renn's historiographical project culminates in what might be termed a "structuralist absolutism," where Einstein becomes the final exhibit in a systematic repudiation of individual agency. Rather than merely contextualizing Einstein's achievements, \emph{The Einsteinian Revolution} suggests that creativity is an illusion subsumed entirely within networks of inherited knowledge, \emph{epistemic matrices}, and conceptual reconfiguration. The result is a historiography that no longer leaves space for exceptional insight, aesthetic daring, or epistemic risk-taking. Gutfreund and Renn construct a counter-myth of structural inevitability in rejecting the myth of genius.

This interpretive posture has ironic consequences. It transforms Beller's Dialogism, a theory conceived originally to recover the situated creativity of scientific actors, into a mechanism for depersonalizing knowledge. 
Gutfreund and Renn preserve only the shell of Beller's theory, stripping it of its animating humanism. Dialogue becomes a neutral exchange between epistemic systems, not a creative struggle among thinkers. The result is a semiotic-cognitive model that processes knowledge flows rather than engages with intellectual personalities. This inversion of Beller's intent is more than a methodological disagreement; it distorts her legacy. Dialogism, conceived to illuminate the emergence of creative thought, risks reducing its original intent.

The central philosophical question raised by \emph{The Einsteinian Revolution} is whether a model of science that excludes individual creativity can meaningfully account for radical conceptual innovation. Gutfreund and Renn implicitly answer negatively: originality becomes synonymous with structural rearrangement. Einstein's theoretical contributions are reframed as inevitable products of epistemic tension rather than visionary acts of reimagining.
This is the paradox of structuralist historiography. In attempting to explain the genesis of ideas through systemic forces, it risks erasing the very agency it seeks to understand. Einstein becomes emblematic not of creative rupture, but of epistemic convergence. His philosophical daring—the rejection of the ether, the relativization of simultaneity, the redefinition of gravity, and the resolution of the Mercury problem—is rendered secondary to the historical processes that supposedly made such moves necessary.

In demythologizing Einstein, Gutfreund and Renn do not merely revise the story—they displace its protagonist. A depersonalized history remains, where a mechanism replaces imagination, and the scientist dissolves into the structure. This, ultimately, is the structuralist dilemma: to explain too much is to explain away.

\section{Conclusion: Restoring Einstein as a Site of Inquiry}

Jürgen Renn’s transition from focused Einstein studies to global narratives like the Anthropocene is not a departure from his structuralist methodology but its logical culmination. His intellectual trajectory has always favored systemic transformation over individual creativity, long-term epistemic shifts over discrete conceptual ruptures. In this framework, Einstein becomes not a source of unresolved philosophical tension or theoretical daring, but a symbolic case study already absorbed into the smooth unfolding of structural reorganization.

Nowhere is this clearer than in \emph{The Einsteinian Revolution}, where Einstein’s conceptual reorientation—from absolute space and time to dynamical spacetime—is recoded as a metaphor for broader shifts in knowledge organization. The planetary thinking required by the Anthropocene—co-evolutionary, interlinked, and systemic—echoes Einstein’s epistemic break. However, Einstein, the thinker, disappeared in being transformed into a heuristic device for global knowledge architecture. His doubts, risk-taking, intellectual independence, and leaps of imagination are replaced by a teleological narrative in which his work is simply one node in a vast cognitive system.

This move has historiographical consequences. As discussed in the section on Mara Beller's dialogism, the flattening of Einstein’s originality is not merely interpretive—it is strategic. Once Einstein is assigned his role within the “Copernican reorganization” of knowledge, there is no need to revisit him—except as a rhetorical touchstone in more sweeping accounts of science, technology, and society. The figure who once unsettled the foundations of physics is now presented as a stabilizing element in the evolution of epistemic order. In Gutfreund and Renn’s account, Einstein is no longer a live question but a closed file.

This also explains the tone of the book. \emph{The Einsteinian Revolution} is not an invitation to renewed debate about Einstein’s creativity, philosophical commitments, or heuristic method. It is an effort to conclude the historiographical chapter—to seal off Einstein’s disruptive potential by embedding it within a system of epistemic constraints and inherited "scaffolding." Despite its title, the book performs a kind of historiographical counter-revolution: it contains the figure of Einstein within a structuralist closure that admits no real agency, only functional transformation.

Yet this closure is not inevitable. If \emph{The Einsteinian Revolution} marks the terminus of a particular structuralist historiography, it also opens space for a critical reappraisal. As argued earlier, the misappropriation of dialogism—as a tool for epistemic smoothing rather than for preserving multiplicity, struggle, and creative rupture—signals the need for its recovery. Beller’s original vision saw scientific knowledge as an inherently dialogical process, irreducible to a single system or trajectory. Her approach foregrounded cognitive conflict, philosophical ambiguity, and the unresolved nature of conceptual innovation.

This article reclaims that dialogical impulse. It resists the transformation of Einstein into a structural cipher and insists on restoring his agency, not as myth but as method—as an enduring site of inquiry. Reintroducing Einstein’s creativity is not to re-mythologize him but to reassert the open-endedness of scientific thought, which is dependent on individuals who cross conceptual boundaries and reshape frameworks from within. Rather than accepting structuralism’s epistemic finality, this article makes a counter-move: it reopens Einstein’s case, not to repeat old hagiographies, but to refuse a historiography that seeks closure where there is still conceptual life.


\begin{thebibliography}{22}

\bibitem [1] {Bel} Beller, M. (1999). \emph{Quantum Dialogues, the Making of a Revolution}. Chicago: University of Chicago Press. 

\bibitem[2] {BRLR1} Blum, A. S., Lalli, R., and Renn, J. (2016). “The Renaissance of General Relativity: How and Why It Happened.” \emph{Annalen der Physik} 528, pp. 344–349. 

\bibitem[3] {BRLR2} Blum, A. S., Lalli, R., and Renn, J. (2018). “Gravitational Waves and the Long Relativity Revolution.” \emph{Nature Astronomy}, pp. 534–543.

\bibitem[4] {Brau} Braudel, F. (1958) “Histoire et Sciences sociales: La longue durée.” Annales 13, pp. 725-753.

\bibitem[5] {CPAE4} \emph{The Collected Papers of Albert Einstein. Vol. 4: The Swiss Years: Writings, 1912–1914}, Klein, M. J., Kox, A. J., Renn, J., and Schulmann, R. (eds.), Princeton: Princeton University Press, 1995. 

\bibitem[6] {Ein14} Einstein, A. (1914). "Die formale Grundlage der allgemeinen Relativitätstheorie." \emph{Königlich Preußische Akademie der Wissenschaften (Berlin). Sitzungsberichte}, pp. 1030-1085.

\bibitem[7] {Ein15-1} Einstein, A. (1915). "Zur allgemeinen Relativitätstheorie." \emph{Königlich Preußische Akademie der Wissenschaften (Berlin). Sitzungsberichte}, pp. 778-786. 

\bibitem[8] {Ein15-2} Einstein, A. (1915). "Erklärung der Perihelbewegung des Merkur aus der allgemeinen Relativitätstheorie." \emph{Königlich Preußische Akademie der Wissenschaften (Berlin). Sitzungsberichte}, pp. 831-839. 

\bibitem[9] {Ein15-3} Einstein, A. (1915). "Die Feldgleichungen der Gravitation", \emph{Königlich Preußische Akademie der Wissenschaften (Berlin). Sitzungsberichte}, pp. 844-847.

\bibitem[10] {EG} Einstein, A. and Grossmann M. (1913). \emph{Entwurf einer verallgemeinerten Relativitätstheorie und einer Theorie der Gravitation I. Physikalischer Teil von Albert Einstein. II. Mathematischer Teil von Marcel Grossmann}. Leipzig and Berlin: B. G. Teubner.

\bibitem[11] {Fl-1} Fleck, L. (1935). \emph{Genesis and Development of a Scientific Fact}. Bradley, F. and Trenn, T. J. (trans.), Chicago: University of Chicago Press, 1979. 

\bibitem[12] {Fl-2} Fleck, L. (1960). “Crisis in Science.” In \emph{Cognition and Fact: Materials on Ludwik Fleck}. Cohen, R. S. and Schnelle, T. (eds.), \emph{Boston Studies in the Philosophy of Science}, vol. 87. Dordrecht: D. Reidel, 1986, pp. 153–158.

\bibitem[13] {Gut} Gutfreund, H. (2019). "The General Theory of Relativity. The Magna Carta of modern physics." In Einstein, A. (1916). \emph{The General Theory of Relativity: Die Grundlage der Allgemeinen Relativitätstheorie}. Cambremer: SP Books, 2019.  

\bibitem[14] {GR1} Gutfreund, H. and Renn, J. (2015). \emph{The Road to Relativity: The History and Meaning of Einstein's "The Foundation of General Relativity."}. Princeton University Press.

\bibitem[15] {GR2} Gutfreund, H. and Renn, J. (2018). \emph{The Formative Years of Relativity: The History and Meaning of Einstein’s Princeton Lectures}. Princeton University Press.

\bibitem[16] {GR3} Gutfreund, H. and Renn, J. (2020). \emph{Einstein on Einstein: Autobiographical and Scientific Reflections}. Princeton University Press.

\bibitem[17] {GR4} Gutfreund, H. and Renn, J. (2024). \emph{The Einsteinian Revolution: The Historical Roots of His Breakthroughs}. Princeton: Princeton University Press. 

\bibitem[18] {Jan} Janssen, M. and Renn, J. (2015). "Arch and scaffold: How Einstein found his field equations." \emph{Physics Today} 68, pp. 30-36.

\bibitem[19] {Kun} Kuhn, T. (1996). \emph{The Structure of Scientific Revolutions}. Chicago: University of Chicago Press.

\bibitem[20] {Lor}  Lorentz, H. A. (1895). \emph{Versuch einer Theorie der elektrischen und optischenen Erscheinungen in bewegten Körpern}. Leiden: E.J. Brill.

\bibitem[21] {Po} Poincaré, H. (1905). "Sur la dynamique de l'électron." \emph{Rendiconti del Circolo Matematico di Palermo} 21, 1906, pp. 1-47. 

\bibitem[22] {RR} Renn, J. and Rynasiewicz, R. (2014). “Einstein’s Copernican Revolution.” In \emph{The Cambridge Companion to Einstein}. Janssen, M. and Lehner, C. (eds.). Cambridge: Cambridge University Press.

\bibitem[23] {Ren} Renn, J. (2020). \emph{The Evolution of Knowledge: Rethinking Science for the Anthropocene}. Princeton: Princeton University Press. 


\end{thebibliography}
\end{document}